\documentclass[12pt,a4paper]{article}
\usepackage{graphicx}

\begin{document}
\title{Static configurations of an exciton field in interaction with a deformable two-dimensional hexagonal lattice}
\date{}
\author{M.J.Watson\footnote{m.j.watson@durham.ac.uk}  and W.J.Zakrzewski\footnote{w.j.zakrzewski@durham.ac.uk}\\
Department of Mathematical Sciences, University of Durham,\\ Durham DH1 3LE, UK}
\maketitle

\begin{abstract}
A model describing the dynamics of crystal exciton states on a hexagonal lattice has been studied in [1]. The hamiltonian includes terms describing the movement of the exciton throughout the crystal. Here we generalize the hamiltonian by including first order couplings of such terms to the lattice phonon fields. A discussion of the ground state configuration of the exciton field as a function of the system couplings is given. Finally we analyse the stability of certain excited states under perturbations.
\end{abstract}

\section{Introduction}

The promotion into the conduction band of an electron in a solid results in the creation of a hole in the valence band, which can be viewed as carrying a positive charge. The electron and hole therefore feel a Coulomb attraction. A quasiparticle may result, known as an exciton, with an energy lower than the bottom of the conduction band. The conduction electron acts to partially screen the interaction between the surrounding nuclei and this results in a modification to the lattice potential. The neighbouring sites are drawn in and this produces a 'self-trapping' of the exciton. 

The importance of this property of the exciton was first realised by Davydov and Kislukha in 1973 [2]. Working in one dimension, they proposed that the exciton state could explain the dispersionless transport of energy in protein chains within the human body - a problem that had been important enough to earn the title: `crisis in bioenergetics' [3]. The quasiparticle moves along the chain, accompanied by its associated lattice deformation. Due to its solitary wave-like properties, the bound state was termed a soliton. The existence of this state was found to exhibit a crucial dependence upon the electron-phonon coupling strength.

The system used in modelling the one-dimensional chain was more recently adapted to describe the two-dimensional case. In the simplest example, i.e. the square lattice, an analysis of the stationary case showed that in the continuum limit the equations governing the evolution of the system could be reduced to a single Nonlinear Schroedinger Equation with an extra term resulting from the discreteness of the lattice [4-6]. This equation, without the extra term, is well known to possess localized solutions in one dimension [7]. In two dimensions however, the soliton is stable only in the presence of the extra term.

The equations were then adapted to the case of a hexagonal lattice [1]. Again soliton solutions were found to exist. Furthermore, the discrete equations were found, in the stationary limit, to obey a Discrete Nonlinear Schroedinger Equation. A similar equation was the starting point for investigations on the hexagonal and triangular lattices [8]. The fact that no such discrete solution could be found in the case of a square lattice suggests that solitons could have an important role in the dynamics of hexagonal crystal lattices, and therefore carbon nanotubes.   

Nanotubes can be thought of geometrically as rolling up a two dimensional hexagonal graphite sheet. The angle at which the sheet is rolled, known as the chiral angle [9], determines whether the carbon nanotube is semi-conducting or metallic. Because of the band properties of semiconductors, it is these types of tubes in which excitons are most important. Due to the nanoscale structure of carbon nanotubes, even a single exciton state could have a considerable effect on the structure's electrical properties.

The system Hamiltonian includes terms that describe the movement of an exciton to a neighbouring lattice site, known as the resonance interaction. Studies so far have worked in the zeroth approximation excluding terms coupling the resonance interaction to the lattice. Here we present the results of our investigations taking into account first order couplings of the resonance interaction to the lattice phonons: terms expected to be important in describing carbon naotubes.  

\section{The Hamiltonian and Equations of Motion}

Our system is described by a Hamiltonian containing three parts, the first part corresponding to the original electron Hamiltonian discussed in [1] (including interaction terms with the lattice), the second to the lattice potential and the third to the new interaction terms introduced in this paper.

\begin{eqnarray}
\hat{H}=\hat{H}_{e}+\hat{H}_{ph}+\hat{H}_{int}
\end{eqnarray}

There are two distinct types of site within the hexagonal lattice, which must be summed over separately in the Hamiltonian. Denoting the distinct types imaginatively as type 1 and type 2 (see figure 1), the Hamiltonian is:

\begin{figure}[ht]\centering
\includegraphics[height=4cm,width=6.5cm,angle=0]{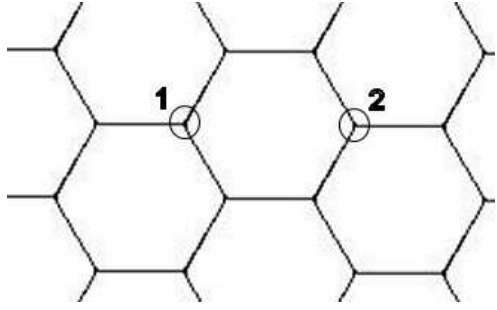}
\caption{\emph{The two inequivalent types of site in the hexagonal lattice.}}
\label{figure 1}
\end{figure}

\begin{eqnarray}
\hat{H}_{e} & = & \sum_{i,j (type1)} 
\biggl[E_0A^{\dagger}_{i,j} A_{i,j}-jA_{i,j}^{\dagger}(A_{i+1,j+1}+A_{i-1,j}+A_{i+1,j-1}) \\
& &  -j(A^{\dagger}_{i+1,j+1}+A^{\dagger}_{i-1,j}+A^{\dagger}_{i+1,j-1})A_{i,j} \nonumber \\
& & +A^{\dagger}_{i,j}A_{i,j}\biggl[\frac{c}{3}(2\hat{u}_{i+1,j}-\hat{u}_{i-1,j+1}-\hat{u}_{i-1,j+1})+\frac{c}{\sqrt{3}}(\hat{v}_{i-1,j+1}-\hat{v}_{i-1,j-1})\biggr] \nonumber \\
& & + \sum_{i,j (type2)} \biggl[E_0A^{\dagger}_{i,j} A_{i,j}-j_{x}A_{i,j}^{\dagger}(A_{i-1,j+1}+A_{i+1,j}+A_{i-1,j-1})\nonumber \\
& &  -j(A^{\dagger}_{i-1,j+1}+A^{\dagger}_{i+1,j}+A^{\dagger}_{i-1,j-1})A_{i,j} \nonumber \\
& & +\sum_{i,j (type2)} A^{\dagger}_{i,j}A_{i,j}\biggl[\frac{c}{3}(2\hat{u}_{i+1,j}-\hat{u}_{i-1,j+1}-\hat{u}_{i-1,j+1})+\frac{c}{\sqrt{3}}(\hat{v}_{i-1,j+1}-\hat{v}_{i-1,j-1})\biggr]\nonumber,
\end{eqnarray}

\begin{eqnarray}
\hat{H_{ph}} & = & \frac{1}{2}\sum_{i,j}\biggl[\biggl(\frac{\hat{p}_{i,j}^{2}}{M}\biggr)+\biggl(\frac{\hat{q}_{i,j}^{2}}{M}\biggr)\biggr]\\
&+&\frac{1}{2}M\sum_{i,j (type 1)}k[(\hat{u}_{i,j}-\hat{u}_{i-1,j})^{2}+(\hat{v}_{i,j}-\hat{v}_{i-1,j})^{2}+(\hat{u}_{i,j}-\hat{u}_{i+1,j+1})^{2}\nonumber \\
&+&(\hat{v}_{i,j}-\hat{v}_{i+1,j+1})^{2}+(\hat{u}_{i,j}-\hat{u}_{i+1,j-1})^{2}+(\hat{v}_{i,j}-\hat{v}_{i+1,j-1})^{2}]\nonumber\\
&+&\frac{1}{2}M\sum_{i,j (type 2)}k[(\hat{u}_{i,j}-\hat{u}_{i+1,j})^{2}+(\hat{v}_{i,j}-\hat{v}_{i+1,j})^{2}+(\hat{u}_{i,j}-\hat{u}_{i-1,j+1})^{2}\nonumber \\
&+&(\hat{v}_{i,j}-\hat{v}_{i-1,j+1})^{2}+(\hat{u}_{i,j}-\hat{u}_{i-1,j-1})^{2}+(\hat{v}_{i,j}-\hat{v}_{i-1,j-1})^{2}]\nonumber,
\end{eqnarray}

\noindent and

\begin{eqnarray}
\hat{H_{int}} & = & -\sum_{i,j (type1)} l\biggl[(A^{\dagger}_{i,j}A_{i+1,j+1}+A^{\dagger}_{i+1,j+1}A_{i,j})\biggl(\frac{1}{3}(\hat{u}_{i+1,j+1}-\hat{u}_{i,j})\\
&+&\frac{1}{\sqrt{3}}(\hat{v}_{i+1,j+1}-\hat{v}_{i,j})\biggr)
+(A^{\dagger}_{i,j}A_{i-1,j}+A^{\dagger}_{i-1,j}A_{i,j})\biggl(\frac{2}{3}(\hat{u}_{i,j}-\hat{u}_{i-1,j})\biggr)\nonumber \\
&+&(A^{\dagger}_{i,j}A_{i+1,j-1}+A^{\dagger}_{i+1,j-1}A_{i,j})\biggl(\frac{1}{3}(\hat{u}_{i+1,j-1}-\hat{u}_{i,j})+\frac{1}{\sqrt{3}}(\hat{v}_{i,j}-\hat{v}_{i+1,j-1})\biggr)\biggr]\nonumber\\
&-& \sum_{i,j (type2)} l\biggl[(A^{\dagger}_{i,j}A_{i-1,j+1}+A^{\dagger}_{i-1,j+1}A_{i,j})\biggl(\frac{1}{3}(\hat{u}_{i,j}-\hat{u}_{i-1,j+1})\nonumber \\
&+&\frac{1}{\sqrt{3}}(\hat{v}_{i-1,j+1}-\hat{v}_{i,j})\biggr)+(A^{\dagger}_{i,j}A_{i+1,j}+A^{\dagger}_{i+1,j}A_{i,j})\biggl(\frac{2}{3}(\hat{u}_{i+1,j}-\hat{u}_{i,j})\biggr)\nonumber \\
&+&(A^{\dagger}_{i,j}A_{i-1,j-1}+A^{\dagger}_{i-1,j-1}A_{i,j})\biggl(\frac{1}{3}(\hat{u}_{i,j}-\hat{u}_{i-1,j-1})+\frac{1}{\sqrt{3}}(\hat{v}_{i,j}-\hat{v}_{i-1,j-1})\biggr)\biggr]\nonumber
\end{eqnarray}

The bosonic creation and annihilation operators for the exciton at site $(i,j)$ are given by $A^{\dagger}_{i,j}$ and $A_{i,j}$ respectively. The operators $\hat{u}_{i,j}$ and $\hat{v}_{i,j}$ give the displacement of molecule $(i,j)$ in the x and y directions, $\hat{p}_{i,j}$ and $\hat{q}_{i,j}$ give their respective conjugated momenta. The zeroth order coupling in the resonance interaction of the exciton is denoted $j$, the new term describing first order couplings of the resonance interaction to the lattice has a strength $l$. The coupling constant $k$ describes the harmonic lattice potential, which, in the limit of small amplitude vibrations, is assumed to provide a good approximation to the true crystal potential. Terms of first order in the displacements in $\hat{H}_{ph}$ are zero due to the associated coupling vanishing in equilibrium. The exciton part of the original Hamiltonian $\hat{H}_{e}$ contains the coupling $c$, which describes the displacement of the lattice in the presence of an exciton. Physically the presence of an exciton at the site $(i,j)$ changes the potential between neighbouring molecules by modifying the interatomic Coulomb force, since an electron has been promoted into a more weakly bound conduction band state. The modification to the Hamiltonian involving the addition of terms proportional to $l$ is given by $\hat{H}_{int}$.

The analysis proceeds in the usual way [4-6] by allowing the Hamiltonian to operate on the approximate product wavefunction:

\begin{eqnarray}
|\Psi(t)\rangle|\Phi(t)\rangle=\sum_{i,j}\psi_{i,j}(t)A^{\dagger}_{i,j}|0\rangle|\Phi(t)\rangle
\end{eqnarray}

\noindent where $|\Psi(t)\rangle$ is an excited state of the lattice and $|\Phi(t)\rangle$ is a coherent phonon state. The states $A^{\dagger}_{i,j}|0\rangle$ are complete and orthonormal. The $\psi_{i,j}$ are the complex coefficients of:

\begin{eqnarray}
\langle\Psi(t)|\Psi(t)\rangle = \sum_{i,j}|\psi_{i,j}(t)|^{2}=1
\end{eqnarray}

\noindent The displacement and momentum operators act upon $|\Phi\rangle$ to give:

\begin{eqnarray}
\langle\Phi|\hat{u}_{i,j}|\Phi\rangle=u_{i,j};~~~~~\langle\Phi|\hat{v}_{i,j}|\Phi\rangle=v_{i,j}\\
\langle\Phi|\hat{p}_{i,j}|\Phi\rangle=p_{i,j};~~~~~\langle\Phi|\hat{q}_{i,j}|\Phi\rangle=q_{i,j}\nonumber
\end{eqnarray}

\noindent The semiclassical Hamiltonian can then be written in terms of the complex amplitudes $\psi_{i,j}(t)$ by calculating:

\begin{eqnarray}
H=\langle\Phi|\langle\Psi|\hat{H}|\Psi\rangle|\Phi\rangle
\end{eqnarray}

\noindent which tranforms the hamiltonian to:

\begin{eqnarray}
H & = & \sum_{i,j (type 1)}
\biggl[(E+W)\psi_{i,j} \psi_{i,j}^{*}-j\psi_{i,j}^{*}(\psi_{i+1,j+1}+\psi_{i-1,j}+\psi_{i+1,j-1})\\
& & {} -j\psi_{i,j}(\psi_{i+1,j+1}^{*}+\psi_{i-1,j}^{*}+\psi_{i+1,j-1}^{*}) \nonumber \\
& & {} + |\psi_{i,j}|^{2}{\biggl( \frac{c}{3}(u_{i+1,j+1}+u_{i+1,j-1}-2u_{i-1,j})+\frac{c}{\sqrt3}(v_{i+1,j+1}-v_{i+1,j-1})\biggr)}\biggr] \nonumber\\
& & -l\biggl[(\psi_{i+1,j+1}^{*}\psi_{i,j}+\psi_{i,j}^{*}\psi_{i+1,j+1})\biggl(\frac{1}{2}(u_{i+1,j+1}-u_{i,j})+\frac{\sqrt{3}}{2}(v_{i+1,j+1}-v_{i,j})\biggr)\nonumber \\ 
& & +(\psi_{i-1,j}^{*}\psi_{i,j}+\psi_{i,j}^{*}\psi_{i-1,j})\biggl(u_{i,j}-u_{i-1,j}\biggr)\nonumber\\
& &+(\psi_{i+1,j-1}^{*}\psi_{i,j}+\psi_{i,j}^{*}\psi_{i+1,j-1})\biggl(\frac{1}{2}(u_{i+1,j-1}-u_{i,j})+\frac{\sqrt{3}}{2}(v_{i,j}-v_{i+1,j-1})\biggr)\biggr]\nonumber\\
& + & \sum_{i,j (type 2)}
\biggl[(E+W)\psi_{i,j} \psi_{i,j}^{*}-j\psi_{i,j}^{*}(\psi_{i+1,j}+\psi_{i-1,j+1}+\psi_{i-1,j-1}) \nonumber \\& & {} -j\psi_{i,j}(\psi_{i+1,j}^{*}+\psi_{i-1,j+1}^{*}+\psi_{i-1,j-1}^{*}) \nonumber \\
& & {} + |\psi_{i,j}|^{2}{\biggl( \frac{c}{3}(-u_{i-1,j-1}-u_{i-1,j+1}+2u_{i+1,j})+\frac{c}{\sqrt3}(v_{i-1,j+1}-v_{i-1,j-1})\biggr)}\biggr] \nonumber\\
& &-l\biggl[(\psi_{i-1,j+1}^{*}\psi_{i,j}+\psi_{i,j}^{*}\psi_{i-1,j+1})\biggl(\frac{1}{2}(u_{i,j}-u_{i-1,j+1})+\frac{\sqrt{3}}{2}(v_{i-1,j+1}-v_{i,j})\biggr)\nonumber \\ 
& & +(\psi_{i+1,j}^{*}\psi_{i,j}+\psi_{i,j}^{*}\psi_{i+1,j})\biggl(u_{i+1,j}-u_{i,j}\biggr)\nonumber\\
& &+(\psi_{i-1,j-1}^{*}\psi_{i,j}+\psi_{i,j}^{*}\psi_{i-1,j-1})\biggl(\frac{1}{2}(u_{i,j}-u_{i-1,j-1})+\frac{\sqrt{3}}{2}(v_{i,j}-v_{i-1,j-1})\biggr)\biggr]\nonumber
\end{eqnarray}

\noindent where the phonon energy $W$ is given by:

\begin{eqnarray}
W & = & \frac{1}{2}M\sum_{j=1}^{N_{2}}\sum_{i=1}^{N_{1}}\biggl[\biggl(\frac{du_{i,j}}{dt}\biggr)^2+\biggl(\frac{dv_{i,j}}{dt}\biggr)^2\biggr]\\
&+&\frac{1}{2}M\sum_{i,j (type2)}(k[(u_{i,j}-u_{i-1,j})^{2}+(v_{i,j}-v_{i-1,j})^{2}+(u_{i,j}-u_{i+1,j+1})^{2}\nonumber \\
&+&(v_{i,j}-v_{i+1,j+1})^{2}+(u_{i,j}-u_{i+1,j-1})^{2}+(v_{i,j}-v_{i+1,j-1})^{2}]\nonumber\\
&+&\frac{1}{2}M\sum_{i,j (type 2)}(k[(u_{i,j}-u_{i+1,j})^{2}+(v_{i,j}-v_{i+1,j})^{2}+(u_{i,j}-u_{i-1,j+1})^{2}\nonumber \\
&+&(v_{i,j}-v_{i-1,j+1})^{2}+(u_{i,j}-u_{i-1,j-1})^{2}+(v_{i,j}-v_{i-1,j-1})^{2}]\nonumber
\end{eqnarray}

The Hamiltonian is thus completely written in terms of the complex field $\psi_{i,j}$ and the actual molecular displacement fields $u_{i,j}$ and $v_{i,j}$. The full equations of motion for the Hamiltonian (9) are given in Appendix A1. Below we give the equations of motion corresponding to the case when the lattice site is of type 1, after performing a set of rescalings to reduce the number of parameters.

\begin{eqnarray}
i\frac{\partial \psi_{i,j}}{\partial \tau}
& = & (E_0 + W_0) \psi_{i,j} - 2(\psi_{i+1,j+1} + \psi_{i-1,j} + \psi_{i+1,j-1}) \\ 
& & +\psi_{i,j} {\biggl[}  ( U_{i+1,j+1} + U_{i+1,j-1}-2U_{i-1,j})+\sqrt{3} (V_{i+1,j+1} - V_{i+1,j-1}){\biggr]}\nonumber \\
& & -L\biggl[\psi_{i+1,j+1}[(U_{i+1,j+1}-U_{i,j})+\sqrt{3}(V_{i+1,j+1}-V_{i,j})] \nonumber\\
& & +2\psi_{i-1,j}[U_{i,j}-U_{i-1,j}] + \psi_{i+1,j-1}[(U_{i+1,j-1}-U_{i,j})+\sqrt{3}(V_{i,j}-V_{i+1,j-1})]\biggr]\nonumber
\end{eqnarray}

\begin{eqnarray}
\frac{d^{2}U_{i,j}}{d\tau^{2}}
& = &-K(3U_{i,j}-U_{i+1,j+1}-U_{i-1,j}-U_{i+1,j-1}) \\
& & - \frac{g}{2}(2|\psi_{i-1,j}|^{2}-|\psi_{i+1,j+1}|^{2}-|\psi_{i+1,j-1}|^{2})+f\biggl[2(\psi_{i-1,j}^{*}\psi_{i,j}+\psi_{i,j}^{*}\psi_{i-1,j})\nonumber\\
& & -(\psi_{i+1,j+1}^{*}\psi_{i,j}+\psi_{i,j}^{*}\psi_{i+1,j+1})-(\psi_{i+1,j-1}^{*}\psi_{i,j}+\psi_{i,j}^{*}\psi_{i+1,j-1})\biggr]\nonumber
\end{eqnarray}

\begin{eqnarray}
\frac{d^{2}V_{i,j}}{dt^{2}}
& = &K(3V_{i,j}-V_{i+1,j+1}-V_{i-1,j}-V_{i+1,j-1}) \\
& &-\frac{\sqrt{3}g}{2}(|\psi_{i+1,j+1}|^{2}-|\psi_{i+1,j-1}|^{2})\nonumber \\
& & +\sqrt{3}f\biggl[(\psi_{i+1,j-1}^{*}\psi_{i,j}+\psi_{i,j}^{*}\psi_{i+1,j-1})-(\psi_{i+1,j+1}^{*}\psi_{i,j}+\psi_{i,j}^{*}\psi_{i+1,j+1})\biggr]\nonumber
\end{eqnarray}  

\noindent where the following rescalings have been performed:
\begin{eqnarray}
& &\tau = \frac{jt}{\hbar}, U = 3Cu, V = 3Cv, E_0=\frac{E}{j}, W_0 =\frac{W}{j},\\
& &L=\frac{2Cl}{j}, C = \frac{c}{9j}, K = \frac{k\hbar^{2}}{j^{2}}, g = \frac{2C^{2}}{E_{s}}, E_s = \frac{Mj}{9\hbar^{2}}, f=\frac{gL}{6C^{2}} .\nonumber
\end{eqnarray}

We have attempted to find solutions to these equations in the stationary limit, both for the discrete case and in the continuum limit. Unfortunately no simple solution was found. In [1] it was found that one could express the discrete four-point laplacian of the relevant combination of the displacement fields in terms of a seven-point laplacian in the exciton field. The hexagonal lattice admitted a simple solution to this equation and this allowed the expression of the system equations of motion in terms of a single discrete nonlinear Schroedinger equation with an extra stabilizing term. The addition of the terms in $l$ makes such a task much more difficult, and as yet no such solution has been found.

\section{Numerical Simulation of the Equations}

Simulation of the equations (11-13) involved the use of a fourth order Runge-Kutta method. In order to reduce the simulation time a compression of the lattice was performed, as discussed in [1]. The Hamiltonian and equations of motion that were directly simulated are presented in Appendix A1. In order to represent carbon nanotubes to a first approximation, continuous boundary conditions were imposed upon the lattice, in both directions. The longitudinal lattice dimension was chosen to be significantly larger than the transverse direction. The lattice size was set to $20$ x $8$ sites. Although this was a relatively small lattice size, it enabled a detailed investiagtion of the system states. A future aim should be to investigate the states on a larger lattice.

\subsection{Finding the Minimum Energy}

\subsubsection{Absorption of Kinetic Energy}
In order to find and then investigate the static configuration through simulation it was necessary to absorb energy from the equations. One method of achieving this involved the addition of the following damping terms to the respective right-hand sides of (4) and (5):

\begin{eqnarray}
-\nu\frac{du_{i,j}}{dt} ~~~~~ \textrm{and}~~~~~  -\nu\frac{dv_{i,j}}{dt},
\end{eqnarray}

\noindent where $\nu$ determines the strength of the damping. These terms have the effect of gradually absorbing the kinetic energy of the molecules vibrating about their equilibrium positions. Figure 2 shows the descent of the system energy for three different values of the damping parameter $\nu$. The parameters used for simulation were $k=1, j=1, l=0$ and $c=23.4$. The initial conditions involved field configurations corresponding to a polaronic like state, i.e. a state localised mainly around a single lattice site. It was known that this state was the minimum energy state for the values of the couplings stated above. The damping was then used to reduce the system energy towards the static configuration. In this particular case it was found that $\nu \approx 1.0$ produced the quickest descent to the minimum energy, however varying the couplings has an effect upon the optimum damping coefficient. Similar effects were observed for other values of $l$ and $c$. 

\begin{figure}[ht]\centering
\includegraphics[height=10cm,width=15cm,angle=0]{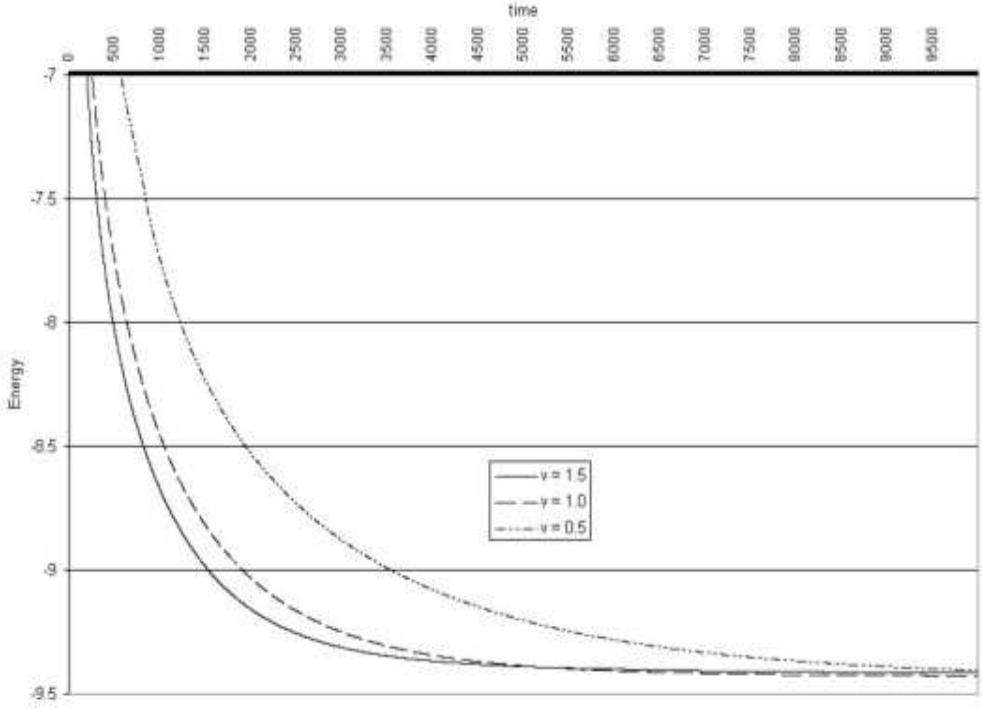}
\caption{\emph{Variation of the system energy due to the addition to the equations of motion of the terms in (15). Three different damping parameters $\nu$ are shown.}}
\label{figure 1}
\end{figure}

\subsubsection{Absorbtion of Energy from the Exciton Field}

The method described in section 3.1.1 was effective in reducing the system energy. However, it was found that the speed of the process could be greatly increased by introducing a direct absorption of energy from the exciton field. The equation of motion for $\psi$ (11) contains only a first order derivative with respect to time, and so the introduction of a damping term similar to that in (15) would destroy the system dynamics. An approximate method was therefore introduced so that the damping term could be included. 

For a quantum state with a definite energy E, the time dependence of the wave function can be taken to be that of a plane wave. The eigenfunctions $\phi(x)$ are then the stationary states of the system:

\begin{eqnarray}
\psi_{i,j}(x,t) = \phi_{i,j}(x)exp\biggl(\frac{-i Et}{\hbar}\biggr)
\end{eqnarray}

\noindent The variation of $\psi$ over an infinitessimal time step t is then:

\begin{eqnarray}
i\hbar \frac{\partial \psi_{i,j}}{\partial t}= E exp\biggl(\frac{-iEt}{\hbar}\biggr)\phi_{i,j} + i\hbar exp\biggl(\frac{-iEt}{\hbar}\biggr)\frac{\partial \phi_{i,j}}{\partial t}
\end{eqnarray}

In the presence of damping, the stationary states $\phi(x,t)$ have some small time dependence.  The value of the energy E was known throughout simulation. The variation of $\psi$ given by (16) could therefore be taken into account. It was then possible to introduce a damping term that would give the variation $\partial_{t}\phi_{i,j}(x,t)$ in (17) without affecting the variation of the stationary states according to (16).

The LHS of (17) is specified by the equation of motion (11). Using (16) to re-write $\psi$ in terms of $\phi$, and cancelling the exponential functions, yields the equation:

\begin{eqnarray}  
i\hbar \frac{\partial \phi_{i,j}}{\partial t}= -E\phi_{i,j} + F(\phi_{i,j}), 
\end{eqnarray}

\noindent where $F(\phi_{i,j})$ is the RHS of equation (11) with $\phi$ replacing $\psi$. The right hand side of this equation was calculated within the simulation. A damping term could then be introduced without destroying the system dynamics:

\begin{eqnarray}
(a+i\hbar) \frac{\partial \phi_{i,j}}{\partial t}= -E\phi_{i,j} + F(\phi_{i,j}) 
\end{eqnarray}

\noindent where the magnitude of $a$ specifies the strength of the absorption.

The implementation of this method resulted in an enormous reduction in the time taken to reach the stationary configuration. It was found that a combination of the two methods for extracting the energy (15) and (19) produced the quickest descent to the minimum energy. Figure 3 shows the variation of the energy in two seperate cases. The first case utilises only the damping given by (15), the second curve uses a combination of the two. The differences are quite astounding.  

\begin{figure}[ht]\centering
\includegraphics[height=9cm,width=13.5cm,angle=0]{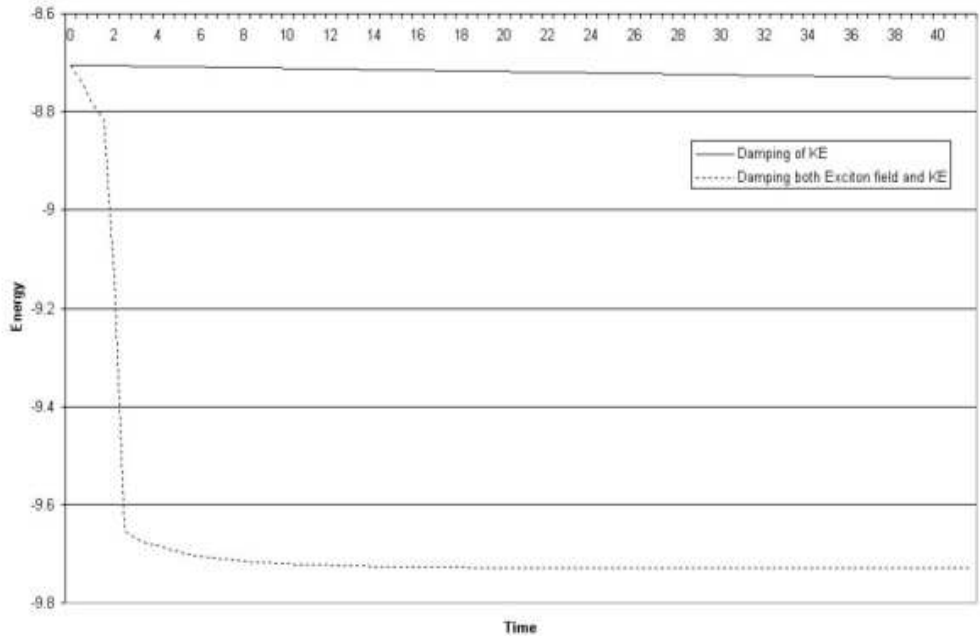}
\caption{\emph{Variation of the system energy due to damping. The first case involves a simple damping of the equations of motion as in (15) with $\nu=1.0$. The second case involves a mixture of damping from the exciton field, and damping of the kinetic energy with $\nu=0.4$. The magnitude of the exciton damping was set proportional to the derivative of the energy with respect to time.}}
\label{figure 2}
\end{figure}

Results showed that the quickest path to the stationary configuration involved making the exciton damping coefficient proportional to the gradient of the energy with respect to time. It was found that the exciton damping becomes less effective when the energy approaches that of the stationary configuration. Under these circumstances the damping of the kinetic energy becomes more useful, particularly for smaller values of $\nu$. Throughout the remainder of the investigation, a value of $\nu=0.4$ was used in damping the system kinetic energy.

\subsection{Minimum energy configurations as a function of c and l}

\subsubsection{Parameters and Method}
The aim was to determine the minimum energy configurations for various values of the couplings $l$ and $c$. The couplings $j$ and $k$ were set equal to 1. Initially the couplings $l$ and $c$ were set to zero, and the minimum energy configuration was found using the methods of the previous section. In fact, for these values of the couplings, the minimum energy configuration corresponded to the completely delocalized state, i.e. where the probability density for the exciton is approximately the same at all lattice sites. The value of $l$ was increased incremently, and the static configuaration was again found. After increasing $l$ upto $14$ the coupling strength was incremently reduced back to zero, following a similar procedure. Upon reaching zero the value of $c$ was increased slightly, and the process repeated upto a maximum value of $c=22$. 

\subsubsection{Results}

Figure 4 shows the maximum in the probability density $|\psi_{i,j}|^{2}_{max}$ for the minimum energy configuration as a function of the couplings $l$ and $c$. The energy corresponding to these configurations is displayed in figure 5.

\begin{figure}[!ht]\centering
\includegraphics[height=10cm,width=15cm,angle=0]{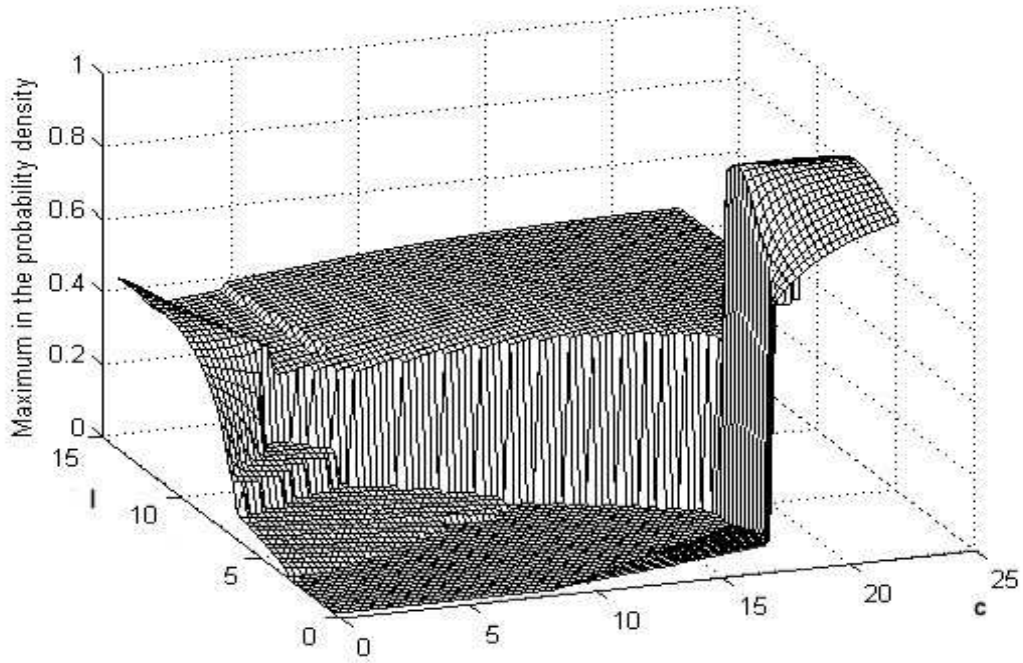}
\caption{\emph{Maximum in the probability density for the minimum energy configuration as a function of the couplings $l$ and $c$.}}
\label{figure 3}
\end{figure}

\begin{figure}[!ht]\centering
\includegraphics[height=10cm,width=15cm,angle=0]{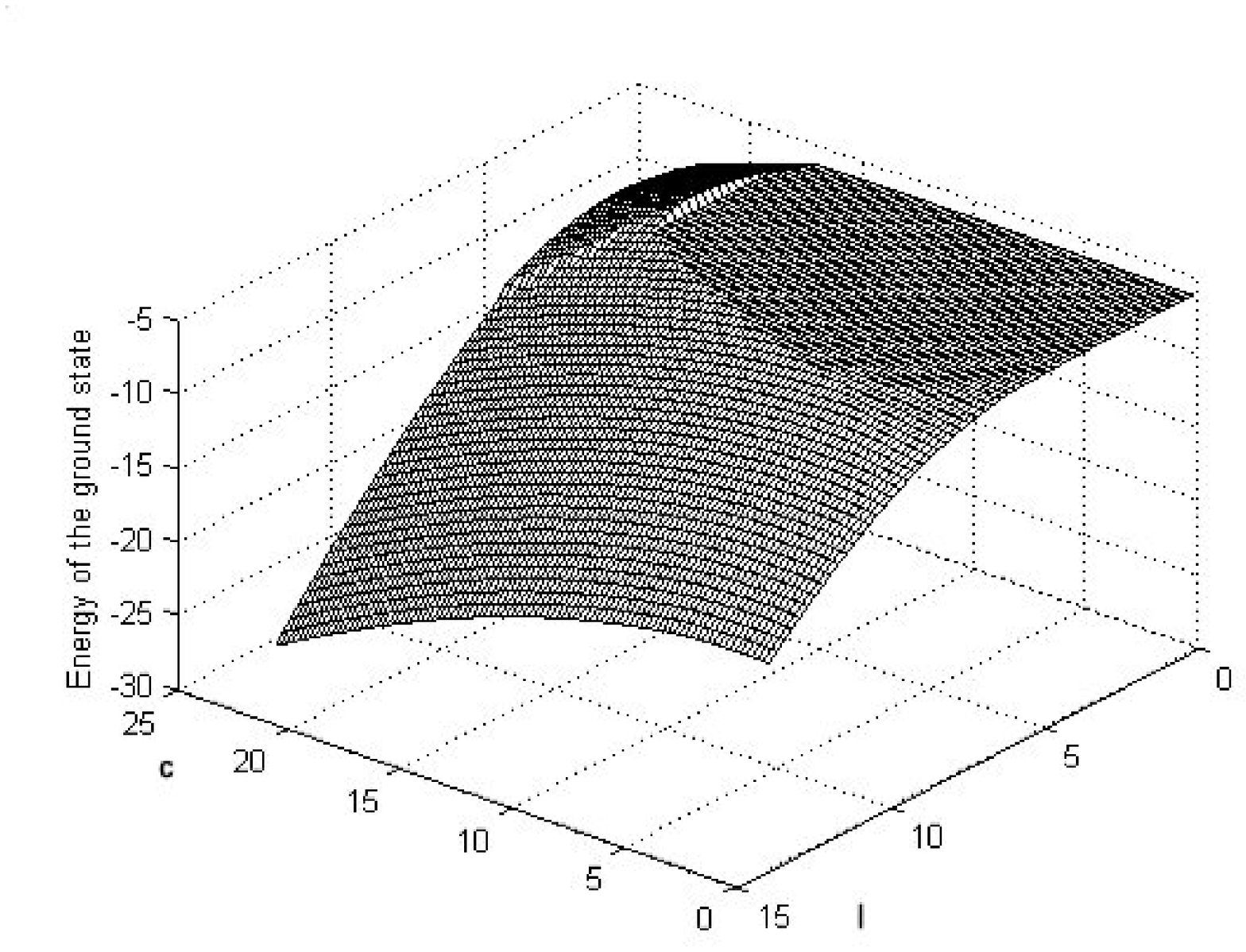}
\caption{\emph{Energy of the ground state configuration as a function of the couplings $l$ and $c$.}}
\label{figure 4}
\end{figure}

As already mentioned, the ground state at $(c,l)=(0,0)$ corresponded to the completely delocalized state. Figure 4 shows how the region where this state corresponds to the ground state extends out to cover a significant proportion of the phase diagram for values of $c$ approximately lying between the limits $0$ and $17$. 

The various field configurations that were observed are shown in figure 6. Keeping $l$ set equal to zero and increasing $c$ had no effect on the configuration of the ground state until $c \approx 8.5$, when the state changed to that shown in figure 6a. Here the probability density for the exciton is localised mainly within a narrow band that wraps around the smallest lattice dimension. Upon increasing $c$ further it was found that this state is stable until $c \approx 18$. For values of $c$ higher than this, the configuration of least energy corresponds to the polaronic type state (fig 6b), localised mainly around a single lattice site. As expected the precise degree of localisation of the polaronic state was found to vary, depending upon the values of the couplings. The maximum in the probability density varied from $0.713$ (c=21, l=3) $< |\psi|^{2}_{max}< 0.998$ (c=17, l=0.25).

\begin{figure}[!ht]\centering
\includegraphics[height=22cm,width=15cm,angle=0]{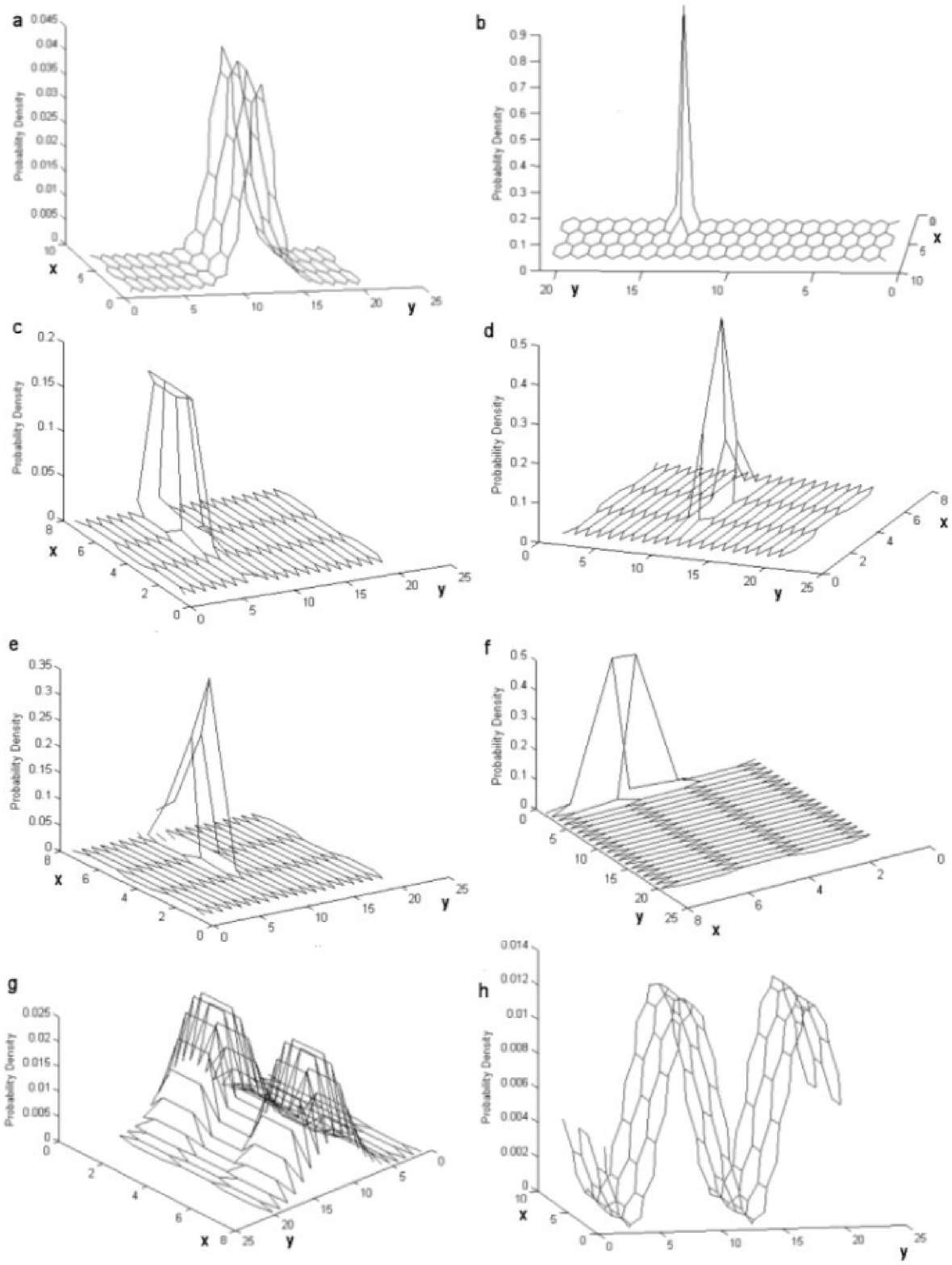}
\caption{\emph{Probability density distribution over the hexagonal lattice of the observed states.}}\label{figure 5}
\end{figure}

Increasing $l$ from the completely delocalized state at $(c,l)=(0,0)$ resulted in an observation of a different set of states. The delocalized state is the ground state until $l \approx 3$. Above this threshold value, the ground state once again corresponded to the wrapped state, already discussed. Increasing $l$ still further resulted in the observation of states localised around six (fig 6c) and then four (fig 6d) sites respectively. Upon increasing $c$ from this area of the phase diagram, the system underwent a smooth transition from that state mainly localised around four sites to a state principally surrounding two lattice sites. Taking $l=13.5$ as an example, as the value of $c$ increases over the region from $1 - 1.5$, the probability density changed smoothly from that shown in fig 6d to that in fig 6f. The two-site state is then the ground state for $c<22$, as is expected to be so for greater values of $c$.

Two other configurations were observed (figure 6 g,h), however the states were always unstable. The state shown in fig 6g shows a two peaked state that is periodic around the nanotube circumference. Fig 6h shows a kind of variation on the singly wrapped state of fig 6a. In this case there exist two separated bands of probability density wrapping the nanotube circumference. 

Each stationary state produced a distinct pattern of lattice displacement. In each case, the lattice sites are drawn in towards the points of high probability density for the exciton. The characteristic pattern of distortion for the polaronic like state is depicted in figure 7a. The displacement of the lattice sites was considerable for those points located close to the centre of the probability distribution. This could have important applications with respect to the electrical properties of hexagonal lattices, and therefore in carbon nanotubes. An extremely interesting case is the wrapped state (fig 6a). The pattern of lattice distortion for this state is given by fig 7b. The transverse displacement (perpendicular to the nanotube axis) is extremely small, approximately of the order $10^{-3}$ the average longitudinal displacement. This state could therefore serve as a starting point for investigation into the effect of excitons upon the electrical properties of carbon nanotubes. The state wraps around the circumference of the nanotube, and so any charge flowing parallel to the nanotube axis encounters approximately the same distortion.

\begin{figure}[!ht]\centering
\includegraphics[height=12cm,width=15cm,angle=0]{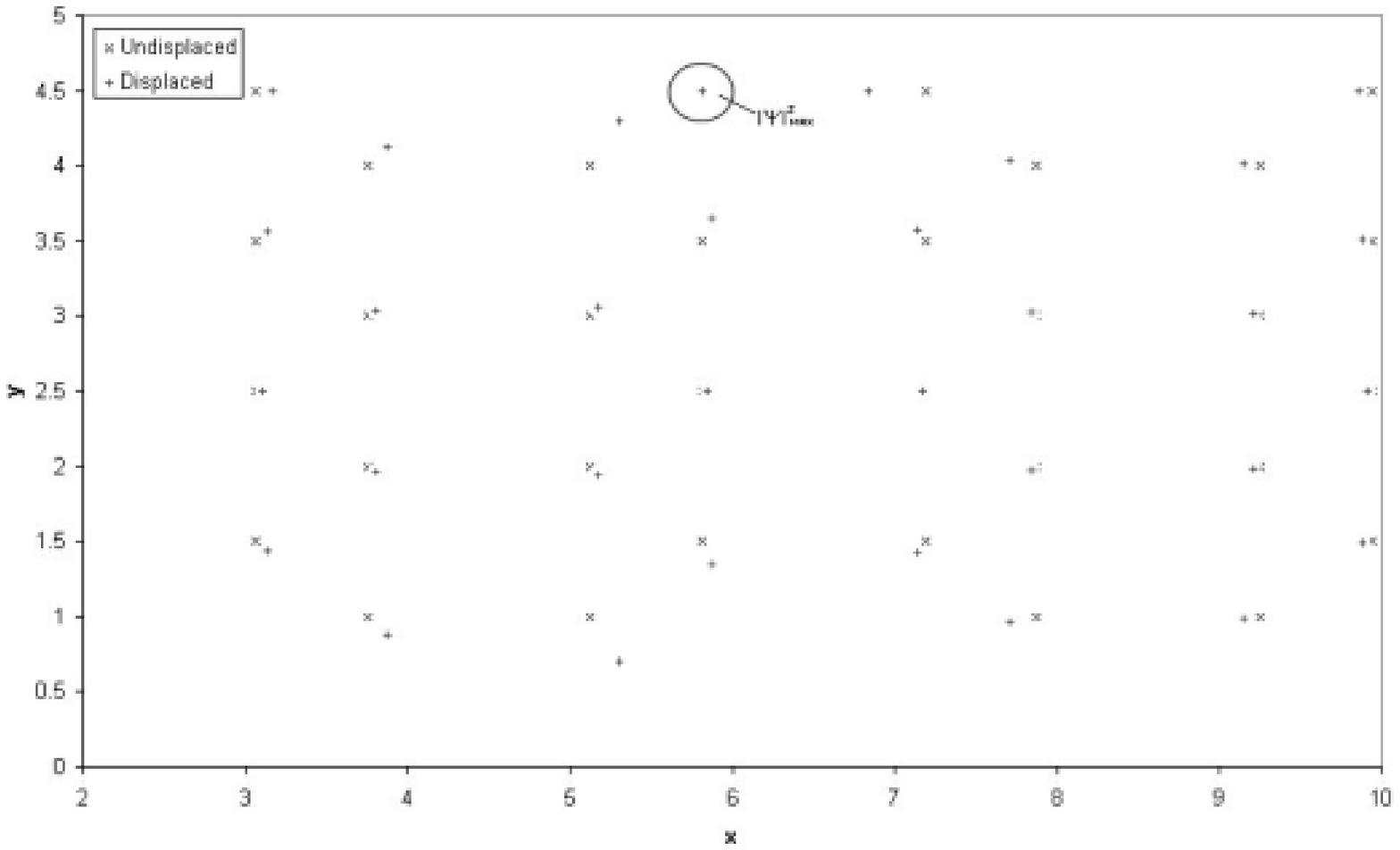}
\caption{\emph{Pattern of lattice displacement centred around a polaronic type state.}}\label{figure 7a}
\end{figure}

\begin{figure}[!ht]\centering
\includegraphics[height=15cm,width=12cm,angle=0]{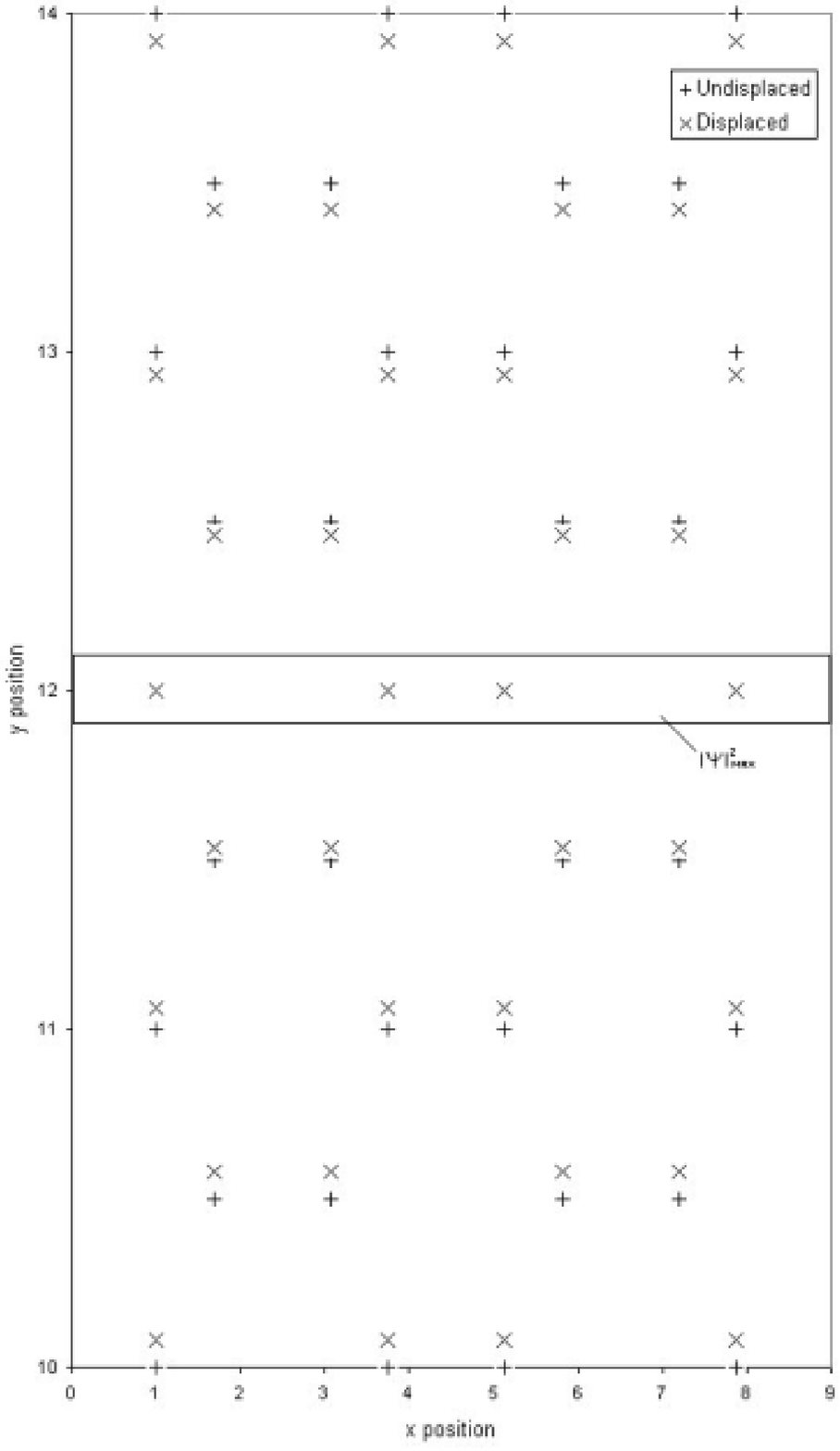}
\caption{\emph{Pattern of lattice displacement for the singly wrapped state.}}\label{figure 7b}
\end{figure}

\subsection{Stability Analysis of States}

The states discussed in the previous section were found to exist over larger areas of the phase diagram than Figure 4 suggests. Of course, in those regions they are higher energy unstable or metastable states. Thus when we perturb them sufficiently the system should fall to the ground state. This section details the results of perturbing states for values of $c$ and $l$ where they are not the minimum energy configuration. 

\subsubsection{Polaronic States}

The method of perturbing the single-site localised polaronic-like states involved increasing the probability density marginally on the three nearest neighbour lattice sites and renormalising the exciton field. In order to quantify the stability of the states, perturbations of increasing strength were applied and the strength of perturbation required to produce a transition was recorded. The perturbation parameter $p$ specifies the strength. A strength of $p=0$ indicates no perturbation, while $p=1$ corresponds to the three nearest neighbour sites being given equal probability density for the exciton as the initial central site. No perturbations with a strength greater than 1 were induced. Figure 9 shows the strength of perturbation required to induce a transition as a function of the couplings $l$ and $c$. For each distinct value of the couplings the initial conditions were given as those corresponding to the closest stable polaronic state. The value of $p$ was then varied from $0$ to $1$ in increments of 0.02. A strength of $p=1$ indicates that the polaron state was stable for all applied perturbations, while a strength of $p=0$ means that the state decayed away naturally without any perturbation. For values of $p$ in between, some degree of perturbation was required to knock the polaron into a lower energy state. This suggests that the state is a local, but not global, minimum in the corresponding region of the coupling phase space. 

\begin{figure}[!ht]\centering
\includegraphics[height=10cm,width=12cm,angle=0]{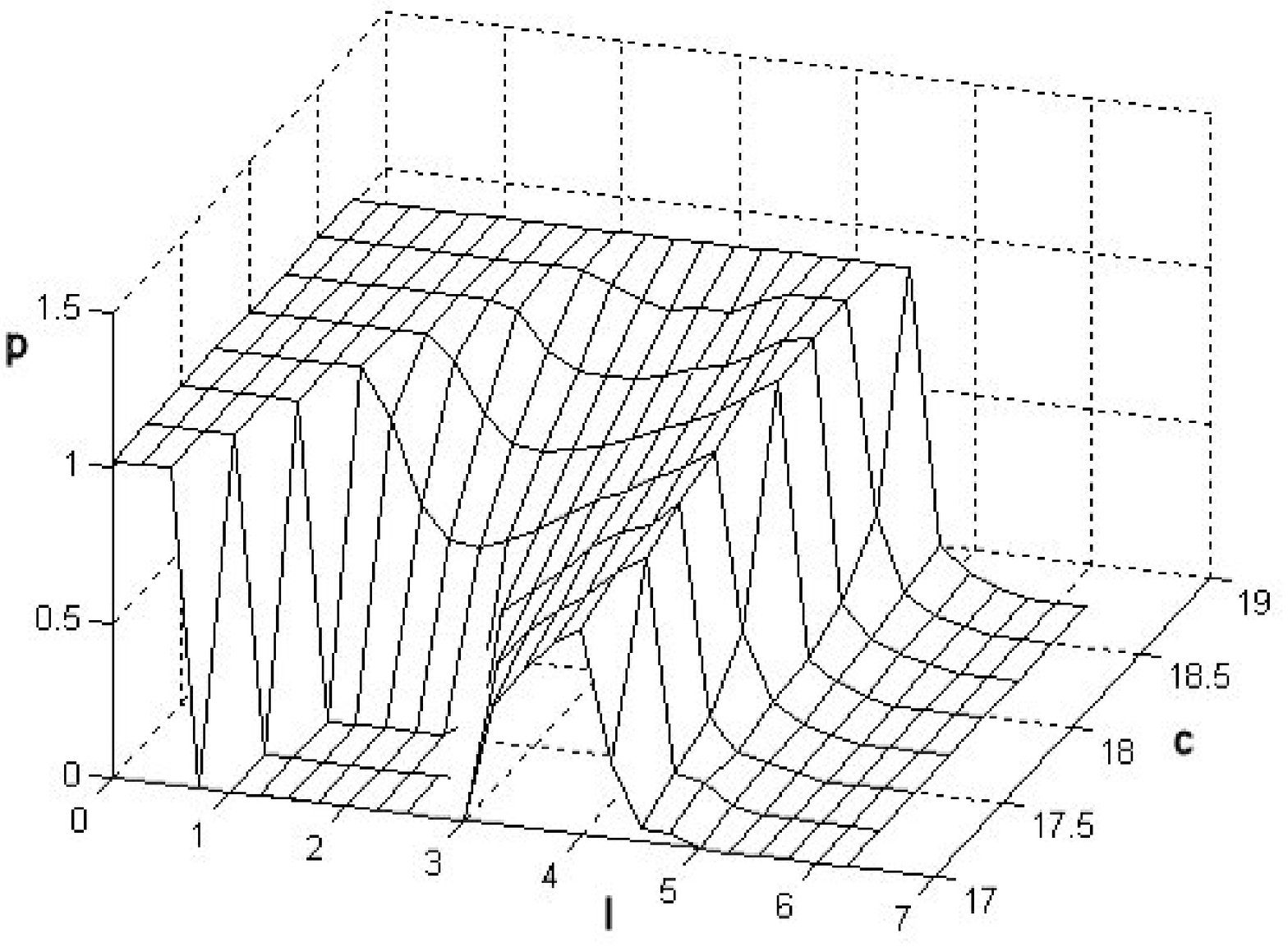}
\caption{\emph{Perturbation strength required to induce a transition from the polaronic state as a function of $c$ and $l$}}\label{figure 8}
\end{figure}

Most values of the couplings displayed by the figure gave either stable (up to $p=1$) polarons or completely unstable polarons. There are are some small regions where metastable states existed. These lie predominantly on the stability boundary between the polaronic states and the two-site localised states. Metastable polaronic states were found to exist for values of $l$ up to and including $l=5.25$. For values of $l$ greater than this the polaron state was found to naturally decay away to the two-site localised state. There exists another region of meta-stability between $3<l<5$. This extends to smaller values of $c<17$. In this region polaron states decayed to the completely delocalised configuration when subject to sufficient perturbation.

\subsubsection{Two-site-localised States}

A similar procedure was undertaken in order to perturb the states localised mainly around two lattice sites (figure 6f). The stability region of these states covers a large proportion of the phase diagram for $c$ and $l$ (figure 4). The initial conditions were those corresponding to the closest two-site localised state. For each value of $c$ the value of $l$ was decreased by different amounts and perturbations were applied, again the aim being to cause a spreading out of the soliton into one of the lower energy states. The perturbations involved increasing the exciton field slightly at the four nearest neighbour sites and renormalising the field. The level of perturbation required to induce a transition was investigated.

The perturbations were in this case ineffective in causing transitions. The states were stable up to perturbations of magnitude $p=1$. However for low enough values of $l$ the state decayed naturally into a lower energy state: either a singly-wrapped state (figure 6a) or the completely delocalised state. Figure 10 shows the maximum in the probability density as $l$ was decreased for different values of $c$. The height of the central peak on the two sites was found to decrease with $l$. To the left of the black line the minimum energy state is that with the exciton field localised mainly around two sites. To the right of this line the minimum energy state corresponded either to the singly-wrapped state or the entirely delocalised state (see figure 4). It is clear from the figure that there exists a region where the states localised around two lattice sites show a large degree of stability, even though the minimum energy configuration was some other state. Furthermore, this region covers a significant proportion of the $l$ versus $c$ phase diagram.

\begin{figure}[!ht]\centering
\includegraphics[height=10cm,width=12cm,angle=0]{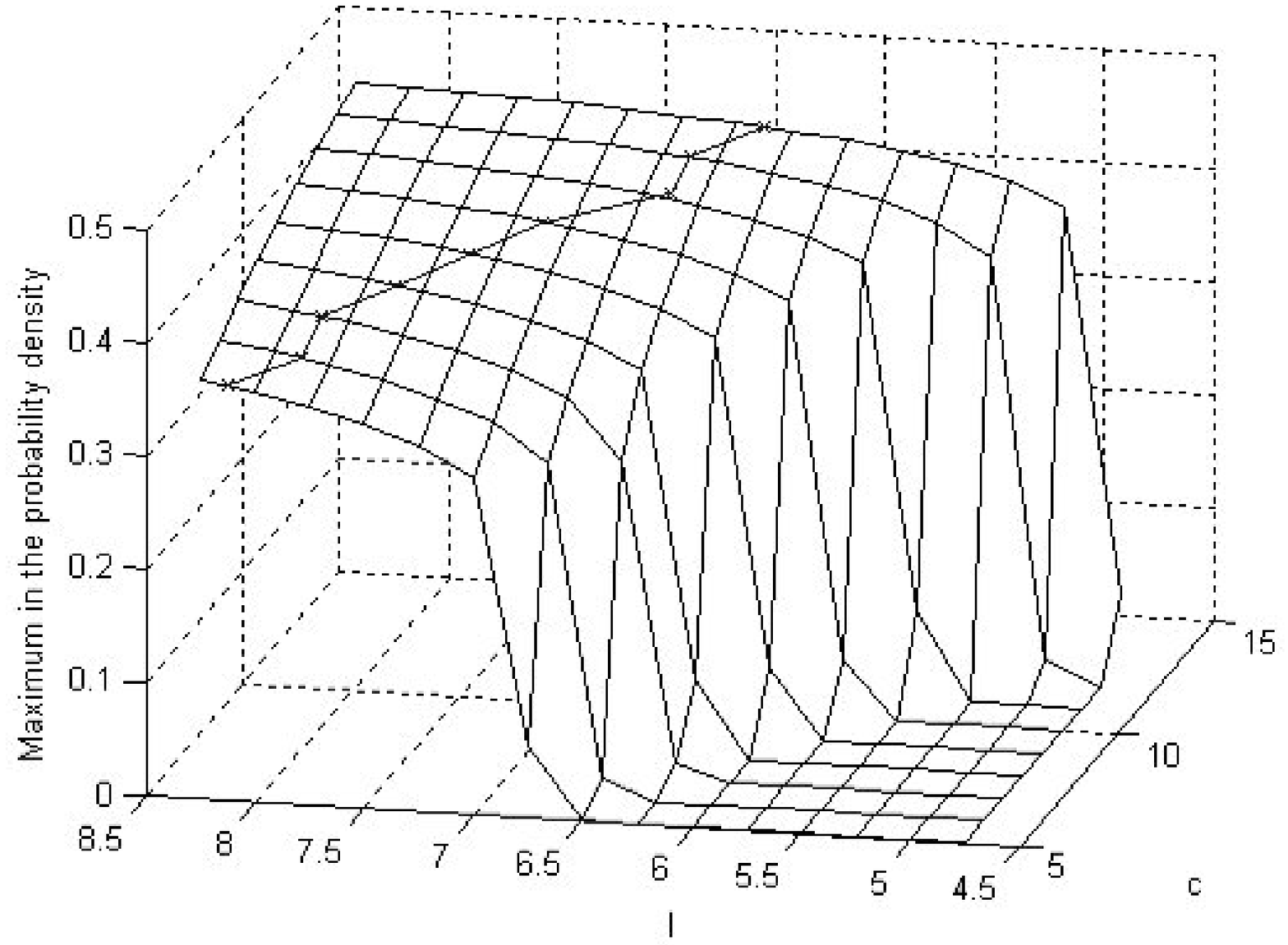}
\caption{\emph{Stability of the two-site-localised state as a function of $c$ and $l$}}\label{figure 9}
\end{figure}

\subsubsection{Stability of other states}

Perturbations were applied to some of the other states displayed in figure 6. The singly-wrapped state (figure 6a) was studied in this way: in particular the boundary between the minimum energy phases of the singly-wrapped state and the completely delocalised state starting at $(c,l)\approx(8.75,0)$. Starting at these values of the couplings, $l$ was gradually increased. The singly-wrapped state was found to decay into the completely delocalised state with extremely little or indeed no perturbation. In each case the transition was extremely gradual, with the maximum in the probability density slowly decreasing with time. This was in stark contrast to the transitions observed from the polaronic and two-site localised states, which were found to decay extremely quickly. The procedure was repeated along the boundary, giving similar results. The state was found to show a large degree of stability only within those regions where it was the minimum energy state. However this is a considerable part of the $c,l$ phase diagram.

The stability of the state localised around six sites was also studied in detail for decreasing $l$. In this case perturbations were applied once again by increasing the magnitude of the exciton field around the soliton in an attempt to encourage it to spread out. Figure 11 displays the results. 

\begin{figure}[!ht]\centering
\includegraphics[height=10cm,width=12cm,angle=0]{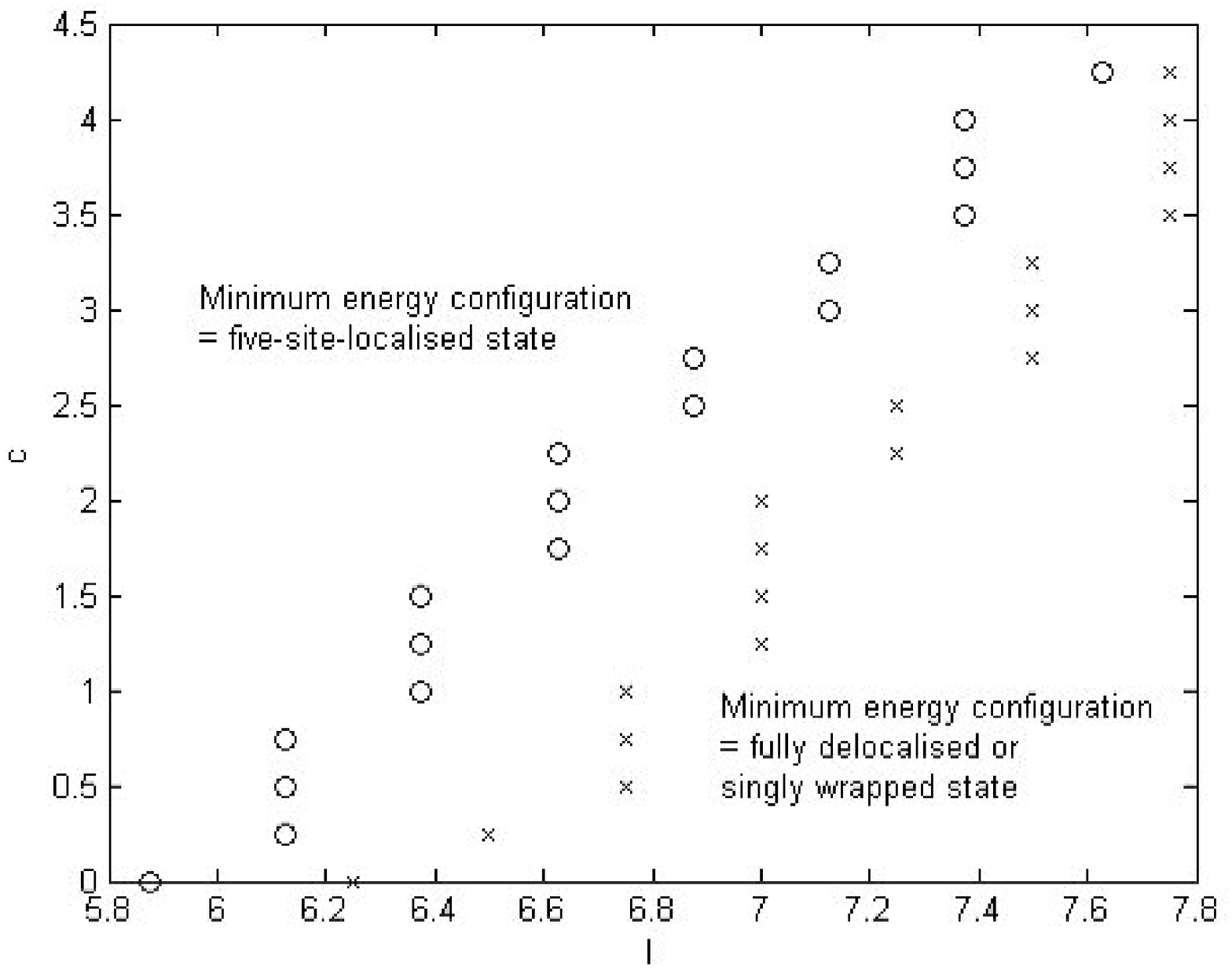}
\caption{\emph{Stability of the six-site-localised state as a function of $c$ and $l$}}\label{figure 10}
\end{figure}

The region to the left of the circles corresponds to the region in figure 4 where the minimum energy state is the six-site-localised state. To the right of the circles the minimum energy configuration is the singly-wrapped state. Perturbations were again applied by increasing the exciton field at the nearest neighbour sites. The narrow band between the lines of circles and crosses in figure 11 is the region where the six-site solitons showed a degree of stability. In fact the states were stable up to values of $p=1$ for all values of the couplings $l$ and $c$ in this region. In the region to the right of the crosses the solitons localised around six sites were found to be unstable, decaying into the singly-wrapped state in the absence of any applied perturbation. 
\section{Conclusion}

Terms describing the coupling of the exciton resonance interaction to lattice displacements were included into a Hamiltonian describing the existence of excitonic states on a hexagonal lattice with continuous boundary conditions. The added terms were found to spoil the elegant analytic solution to the equations of motion found in [1], however numerical simulation showed that these terms produced a rich structure in the coupling phase space of stationary states. By damping the displacement and exciton fields, solitonic stationary states localised around varying numbers of lattice sites were observed including, in agreement with previous results, extremely localised polaron states. Certain solitons were also found to wrap around the lattice, producing patterns of lattice displacement that may perhaps be used to study the effects of solitons upon the electrical properties of carbon nanotubes. The stability of states for certain values of $c$ and $l$ was investigated numerically by introducing perturbations. In most cases the system possessed metastable states in addition to the ground state. Some of these metastable states appeared to be separated from the ground state by a small energy barrier and were thus local minima of the energy. Other excited states were found to exhibit stability under perturbation.

\section{References}

[1] B. Hartmann and W.J. Zakrzewski, {\it Electrons on Hexagonal Lattices and applications to nanotubes}, Phys.Rev. {\bf B 68} (2003) 184302

\noindent[2] A.S.Davydov and N.I. Kislukha  1973, Phys. Stat. Sol. (b) 59,456

\noindent[3] A.S. Davydov, {\it Solitons in Molecular Systems 2nd ed}.,Kluwer Academic (1991)

\noindent[4] L.Brizhik, A. Eremko, B. Piette and W.J. Zakrzewski, Physica {\bf D 146 }(2000), 275.

\noindent[5] L.Brizhik, B. Piette and W.J. Zakrzewski, Ukr. Fiz. Journal {\bf 46 }(2001), 503.

\noindent[6] L.Brizhik, A. Eremko, B. Piette and W.J. Zakrzewski, Physica {\bf D 159 }(2001), 71.

\noindent[7] J.C. Eilbeck and M. Johansson, {\it The Discrete Nonlinear Scroedinger Equation - 20 years on}, arXiv:nlin/0211049

\noindent[8] P.G. Kevrekidis, B.A. Malomed and Yu. B. Gaididei, {\it Solitons in triangular and honeycomb dynamical lattices with the cubic nonlinearity}, Phys. Rev. {\bf E 66} (2002) 016609

\noindent[9] R. Saito, G. Dresselhaus and M. S. Dresselhaus, {\it Physical properties of carbon nanotubes}, World Scientific (1998)

\section{Appendix A1 - Hamiltonian and equations of motion for simulation}

The simulated equations are altered to account for four different lattice sites separately. The advantage of this method lies in the reduction of the array sizes used throughout the simulations. For more information see [1]. The Hamiltonian used throughout was:

\begin{eqnarray}
H & = & \sum_{\frac{j-1}{2}=0}^{\frac{N_{2}}{2}-1} \sum_{\frac{i-1}{4}=0}^{\frac{N_{1}}{2}-3}
\biggl[(E+W)\psi_{i,j} \psi_{i,j}^{*}-j\psi_{i,j}^{*}(\psi_{i+1,j}+\psi_{i-1,j}+\psi_{i+1,j-1})\\
& & {} -j\psi_{i,j}(\psi_{i+1,j}^{*}+\psi_{i-1,j}^{*}+\psi_{i+1,j-1}^{*}) \nonumber \\
& & {} + |\psi_{i,j}|^{2}{\biggl( \frac{c}{3}(u_{i+1,j}+u_{i+1,j-1}-2u_{i-1,j})+\frac{c}{\sqrt3}(v_{i+1,j}-v_{i+1,j-1})\biggr)}\biggr] \nonumber\\
& & -l\biggl[(\psi_{i+1,j}^{*}\psi_{i,j}+\psi_{i,j}^{*}\psi_{i+1,j})\biggl(\frac{1}{2}(u_{i+1,j}-u_{i,j})+\frac{\sqrt{3}}{2}(v_{i+1,j}-v_{i,j})\biggr)\nonumber \\ 
& & +(\psi_{i-1,j}^{*}\psi_{i,j}+\psi_{i,j}^{*}\psi_{i-1,j})\biggl(u_{i,j}-u_{i-1,j}\biggr)\nonumber\\
& &+(\psi_{i+1,j-1}^{*}\psi_{i,j}+\psi_{i,j}^{*}\psi_{i+1,j-1})\biggl(\frac{1}{2}(u_{i+1,j-1}-u_{i,j})+\frac{\sqrt{3}}{2}(v_{i,j}-v_{i+1,j-1})\biggr)\biggr]\nonumber\\
& + & \sum_{\frac{j}{2}=1}^{\frac{N_{2}}{2}} \sum_{\frac{i-2}{4}=0}^{\frac{N_{1}}{4}-2}
\biggl[(E+W)\psi_{i,j} \psi_{i,j}^{*}-j\psi_{i,j}^{*}(\psi_{i+1,j}+\psi_{i-1,j+1}+\psi_{i-1,j}) \nonumber \\
& & {} -j\psi_{i,j}(\psi_{i+1,j}^{*}+\psi_{i-1,j+1}^{*}+\psi_{i-1,j}^{*}) \nonumber \\
& & {} + |\psi_{i,j}|^{2}{\biggl( \frac{c}{3}(-u_{i-1,j}-u_{i-1,j+1}+2u_{i+1,j})+\frac{c}{\sqrt3}(v_{i-1,j+1}-v_{i-1,j})\biggr)}\biggr] \nonumber\\
& &-l\biggl[(\psi_{i-1,j+1}^{*}\psi_{i,j}+\psi_{i,j}^{*}\psi_{i-1,j+1})\biggl(\frac{1}{2}(u_{i,j}-u_{i-1,j+1})+\frac{\sqrt{3}}{2}(v_{i-1,j+1}-v_{i,j})\biggr)\nonumber \\ 
& & +(\psi_{i+1,j}^{*}\psi_{i,j}+\psi_{i,j}^{*}\psi_{i+1,j})\biggl(u_{i+1,j}-u_{i,j}\biggr)\nonumber\\
& &+(\psi_{i-1,j}^{*}\psi_{i,j}+\psi_{i,j}^{*}\psi_{i-1,j})\biggl(\frac{1}{2}(u_{i,j}-u_{i-1,j})+\frac{\sqrt{3}}{2}(v_{i,j}-v_{i-1,j})\biggr)\biggr]\nonumber\\
& + &  \sum_{\frac{j}{2}=1}^{\frac{N_{2}}{2}} \sum_{\frac{i-3}{4}=0}^{\frac{N_{1}}{4}-1}
\biggl[(E+W)\psi_{i,j} \psi_{i,j}^{*}-j\psi_{i,j}^{*}(\psi_{i+1,j+1}+\psi_{i-1,j}+\psi_{i+1,j})\nonumber\\
& & {} -j\psi_{i,j}(\psi_{i+1,j+1}^{*}+\psi_{i-1,j}^{*}+\psi_{i+1,j}^{*}) \nonumber \\
& & {} + |\psi_{i,j}|^{2}{\biggl( \frac{c}{3}(u_{i+1,j+1}+u_{i+1,j}-2u_{i-1,j})+\frac{c}{\sqrt3}(v_{i+1,j+1}-v_{i+1,j})\biggr)}\biggr] \nonumber\\
& & -l\biggl[(\psi_{i+1,j+1}^{*}\psi_{i,j}+\psi_{i,j}^{*}\psi_{i+1,j+1})\biggl(\frac{1}{2}(u_{i+1,j+1}-u_{i,j})+\frac{\sqrt{3}}{2}(v_{i+1,j+1}-v_{i,j})\biggr)\nonumber \\ 
& & +(\psi_{i-1,j}^{*}\psi_{i,j}+\psi_{i,j}^{*}\psi_{i-1,j})\biggl(u_{i,j}-u_{i-1,j}\biggr)\nonumber\\
& &+(\psi_{i+1,j}^{*}\psi_{i,j}+\psi_{i,j}^{*}\psi_{i+1,j})\biggl(\frac{1}{2}(u_{i+1,j}-u_{i,j})+\frac{\sqrt{3}}{2}(v_{i,j}-v_{i+1,j})\biggr)\biggr]\nonumber\\
& + & \sum_{\frac{j-1}{2}=0}^{\frac{N_{2}}{2}-1} \sum_{\frac{i}{4}=1}^{\frac{N_{1}}{4}}
\biggl[(E+W)\psi_{i,j} \psi_{i,j}^{*}-j\psi_{i,j}^{*}(\psi_{i+1,j}+\psi_{i-1,j}+\psi_{i-1,j-1}) \nonumber \\
& & {} -j\psi_{i,j}(\psi_{i+1,j}^{*}+\psi_{i-1,j}^{*}+\psi_{i-1,j-1}^{*}) \nonumber \\
& & {} + |\psi_{i,j}|^{2}{\biggl( \frac{c}{3}(-u_{i-1,j-1}-u_{i-1,j}+2u_{i+1,j})+\frac{c}{\sqrt3}(v_{i-1,j}-v_{i-1,j-1})\biggr)}\biggr] \nonumber\\
& &-l\biggl[(\psi_{i-1,j}^{*}\psi_{i,j}+\psi_{i,j}^{*}\psi_{i-1,j})\biggl(\frac{1}{2}(u_{i,j}-u_{i-1,j})+\frac{\sqrt{3}}{2}(v_{i-1,j}-v_{i,j})\biggr)\nonumber \\ 
& & +(\psi_{i+1,j}^{*}\psi_{i,j}+\psi_{i,j}^{*}\psi_{i+1,j})\biggl(u_{i+1,j}-u_{i,j}\biggr)\nonumber\\
& &+(\psi_{i-1,j-1}^{*}\psi_{i,j}+\psi_{i,j}^{*}\psi_{i-1,j-1})\biggl(\frac{1}{2}(u_{i,j}-u_{i-1,j-1})+\frac{\sqrt{3}}{2}(v_{i,j}-v_{i-1,j-1})\biggr)\biggr]\nonumber
\end{eqnarray}

The corresponding equations of motion are, for $\psi$:

$i=1+4k$
\begin{eqnarray} 
i\hbar \frac{\partial \psi_{i,j}}{\partial t} 
 & = & (E + W) \psi_{i,j} - 2j (\psi_{i+1,j} + \psi_{i-1,j} + \psi_{i+1,j-1}) \nonumber \\ 
& & {}+\psi_{i,j} {\displaystyle\biggl[} \frac{c_x}{3} ( u_{i+1,j} + u_{i+1,j-1}-2u_{i-1,j}) \\
& & {}+\frac{c}{\sqrt 3} (v_{i+1,j} - v_{i+1,j-1}){\biggr]} \nonumber\\
& &-l\biggl[\psi_{i+1,j}[(u_{i+1,j}-u_{i,j})+\sqrt{3}(v_{i+1,j}-v_{i,j})] \nonumber\\
& &+2\psi_{i-1,j}[u_{i,j}-u_{i-1,j}] + \psi_{i+1,j-1}[(u_{i+1,j-1}-u_{i,j})+\sqrt{3}(v_{i,j}-v_{i+1,j-1})]\biggr]\nonumber \end{eqnarray}
$i=2+4k$
\begin{eqnarray} 
i\hbar \frac{\partial \psi_{i,j}}{\partial t} 
 & = & (E + W) \psi_{i,j} - 2j (\psi_{i+1,j} + \psi_{i-1,j} + \psi_{i-1,j+1}) \nonumber \\ 
& & {}+\psi_{i,j} {\displaystyle\biggl[} \frac{c_x}{3} (2u_{i+1,j} - u_{i-1,j+1}-u_{i-1,j}) \\
& & {}+\frac{c}{\sqrt 3} (v_{i-1,j+1} - v_{i-1,j}){\biggr]} \nonumber\\
& &-l\biggl[\psi_{i-1,j+1}[(u_{i,j}-u_{i-1,j+1})+\sqrt{3}(v_{i-1,j+1}-v_{i,j})] \nonumber\\
& &+2\psi_{i+1,j}[u_{i+1,j}-u_{i,j}] + \psi_{i-1,j}[(u_{i,j}-u_{i-1,j})+\sqrt{3}(v_{i,j}-v_{i-1,j})]\biggr]\nonumber 
\end{eqnarray}
$i=3+4k$
\begin{eqnarray} 
i\hbar \frac{\partial \psi_{i,j}}{\partial t} 
 & = & (E + W) \psi_{i,j} - 2j (\psi_{i+1,j} + \psi_{i-1,j} + \psi_{i+1,j+1}) \nonumber \\ 
& & {}+\psi_{i,j} {\displaystyle\biggl[} \frac{c_x}{3} ( u_{i+1,j} + u_{i+1,j+1}-2u_{i-1,j}) \\
& & {}+\frac{c}{\sqrt 3} (v_{i+1,j} - v_{i+1,j-1}){\biggr]} \nonumber\\
& &-l\biggl[\psi_{i+1,j}[(u_{i+1,j}-u_{i,j})+\sqrt{3}(v_{i,j}-v_{i+1,j})] \nonumber\\
& &+2\psi_{i-1,j}[u_{i,j}-u_{i-1,j}] + \psi_{i+1,j+1}[(u_{i+1,j+1}-u_{i,j})+\sqrt{3}(v_{i+1,j+1}-v_{i,j})]\biggr]\nonumber \end{eqnarray}
$i=4+4k$
\begin{eqnarray} 
i\hbar \frac{\partial \psi_{i,j}}{\partial t} 
 & = & (E + W) \psi_{i,j} - 2j (\psi_{i+1,j} + \psi_{i-1,j} + \psi_{i-1,j-1}) \nonumber \\ 
& & {}+\psi_{i,j} {\displaystyle\biggl[} \frac{c_x}{3} (2u_{i+1,j} - u_{i-1,j-1}-u_{i-1,j}) \\
& & {}+\frac{c}{\sqrt 3} (v_{i-1,j} - v_{i-1,j-1}){\biggr]} \nonumber\\
& &-l\biggl[\psi_{i-1,j-1}[(u_{i,j}-u_{i-1,j-1})+\sqrt{3}(v_{i,j}-v_{i-1,j-1})] \nonumber\\
& &+2\psi_{i+1,j}[u_{i+1,j}-u_{i,j}] + \psi_{i-1,j}[(u_{i,j}-u_{i-1,j})+\sqrt{3}(v_{i-1,j}-v_{i,j})]\biggr]\nonumber 
\end{eqnarray}

while equations describing the lattice site displacements $u_{i,j}$ and $v_{i,j}$ are: 

$i=1+4k$
\begin{eqnarray}
\frac{d^{2}u_{i,j}}{dt^{2}}
& = & k(3u_{i,j}-u_{i+1,j}-u_{i-1,j}-u_{i+1,j-1}) \nonumber\\
& & {}+ \frac{c}{3M}(2|\psi_{i-1,j}|^{2}-|\psi_{i+1,j}|^{2}-|\psi_{i+1,j-1}|^{2})\nonumber\\
& & -l\biggl[-(\psi_{i+1,j}^{*}\psi_{i,j}+\psi_{i,j}^{*}\psi_{i+1,j})\nonumber\\
& & +2(\psi_{i-1,j}^{*}\psi_{i,j}+\psi_{i,j}^{*}\psi_{i-1,j}) \nonumber\\
& & -(\psi_{i+1,j-1}^{*}\psi_{i,j}+\psi_{i,j}^{*}\psi_{i+1,j-1})\biggr] 
\end{eqnarray}

$i=2+4k$
\begin{eqnarray}
\frac{d^{2}u_{i,j}}{dt^{2}}
& = & k(3u_{i,j}-u_{i+1,j}-u_{i-1,j}-u_{i-1,j+1}) \nonumber\\
& & {}- \frac{c}{3M}(2|\psi_{i+1,j}|^{2}-|\psi_{i-1,j+1}|^{2}-|\psi_{i-1,j}|^{2})\nonumber\\
& & +l\biggl[-(\psi_{i-1,j+1}^{*}\psi_{i,j}+\psi_{i,j}^{*}\psi_{i-1,j+1})\nonumber\\
& & +2(\psi_{i+1,j}^{*}\psi_{i,j}+\psi_{i,j}^{*}\psi_{i+1,j}) \nonumber\\
& & -(\psi_{i-1,j}^{*}\psi_{i,j}+\psi_{i,j}^{*}\psi_{i-1,j})\biggr] 
\end{eqnarray}

$i=3+4k$
\begin{eqnarray}
\frac{d^{2}u_{i,j}}{dt^{2}}
& = & k(3u_{i,j}-u_{i+1,j}-u_{i-1,j}-u_{i+1,j+1}) \nonumber\\
& & {}+ \frac{c}{3M}(2|\psi_{i-1,j}|^{2}-|\psi_{i+1,j}|^{2}-|\psi_{i+1,j+1}|^{2})\nonumber\\
& & -l\biggl[-(\psi_{i+1,j}^{*}\psi_{i,j}+\psi_{i,j}^{*}\psi_{i+1,j})\nonumber\\
& & +2(\psi_{i-1,j}^{*}\psi_{i,j}+\psi_{i,j}^{*}\psi_{i-1,j}) \nonumber\\
& & -(\psi_{i+1,j+1}^{*}\psi_{i,j}+\psi_{i,j}^{*}\psi_{i+1,j+1})\biggr] 
\end{eqnarray}

$i=4+4k$
\begin{eqnarray}
\frac{d^{2}u_{i,j}}{dt^{2}}
& = & k(3u_{i,j}-u_{i+1,j}-u_{i-1,j}-u_{i-1,j-1}) \nonumber\\
& & {}- \frac{c}{3M}(2|\psi_{i+1,j}|^{2}-|\psi_{i-1,j+1}|^{2}-|\psi_{i-1,j}|^{2})\nonumber\\
& & +l\biggl[-(\psi_{i-1,j-1}^{*}\psi_{i,j}+\psi_{i,j}^{*}\psi_{i-1,j-1})\nonumber\\
& & +2(\psi_{i+1,j}^{*}\psi_{i,j}+\psi_{i,j}^{*}\psi_{i+1,j}) \nonumber\\
& & -(\psi_{i-1,j}^{*}\psi_{i,j}+\psi_{i,j}^{*}\psi_{i-1,j})\biggr] 
\end{eqnarray}

and
$i=1+4k$
\begin{eqnarray}
\frac{d^{2}v_{i,j}}{dt^{2}}
& = & k(3v_{i,j}-v_{i+1,j}-v_{i-1,j}-v_{i+1,j-1}) \nonumber \\
& & {}- \frac{c}{\sqrt 3M}(|\psi_{i+1,j}|^{2}-|\psi_{i+1,j-1}|^{2})\nonumber \\
& & -l\sqrt{3}\biggl[-(\psi_{i+1,j}^{*}\psi_{i,j}+\psi_{i,j}^{*}\psi_{i+1,j})\nonumber\\
& & +(\psi_{i+1,j-1}^{*}\psi_{i,j}+\psi_{i,j}^{*}\psi_{i+1,j-1})\biggr].
\end{eqnarray}
$i=2+4k$
\begin{eqnarray}
\frac{d^{2}v_{i,j}}{dt^{2}}
& = & k(3v_{i,j}-v_{i+1,j}-v_{i-1,j}-v_{i-1,j+1}) \nonumber \\
& & {}- \frac{c}{\sqrt 3M}(|\psi_{i-1,j}|^{2}-|\psi_{i-1,j+1}|^{2})\nonumber \\
& & +l\sqrt{3}\biggl[-(\psi_{i+1,j}^{*}\psi_{i,j}+\psi_{i,j}^{*}\psi_{i+1,j})\nonumber\\
& & +(\psi_{i-1,j+1}^{*}\psi_{i,j}+\psi_{i,j}^{*}\psi_{i-1,j+1})\biggr].
\end{eqnarray}
$i=3+4k$
\begin{eqnarray}
\frac{d^{2}v_{i,j}}{dt^{2}}
& = & k(3v_{i,j}-v_{i+1,j}-v_{i-1,j}-v_{i+1,j+1}) \nonumber \\
& & {}- \frac{c}{\sqrt 3M}(|\psi_{i+1,j}|^{2}-|\psi_{i+1,j+1}|^{2})\nonumber \\
& & +l\sqrt{3}\biggl[-(\psi_{i+1,j}^{*}\psi_{i,j}+\psi_{i,j}^{*}\psi_{i+1,j})\nonumber\\
& & +(\psi_{i+1,j+1}^{*}\psi_{i,j}+\psi_{i,j}^{*}\psi_{i+1,j+1})\biggr].
\end{eqnarray}
$i=4+4k$
\begin{eqnarray}
\frac{d^{2}v_{i,j}}{dt^{2}}
& = & k(3v_{i,j}-v_{i+1,j}-v_{i-1,j}-v_{i-1,j-1}) \nonumber \\
& & {}- \frac{c}{\sqrt 3M}(|\psi_{i-1,j}|^{2}-|\psi_{i-1,j-1}|^{2})\nonumber \\
& & -l\sqrt{3}\biggl[-(\psi_{i+1,j}^{*}\psi_{i,j}+\psi_{i,j}^{*}\psi_{i+1,j})\nonumber\\
& & +(\psi_{i-1,j-1}^{*}\psi_{i,j}+\psi_{i,j}^{*}\psi_{i-1,j-1})\biggr].
\end{eqnarray}

\end{document}